\documentclass[prl,twocolumn]{revtex4}
\usepackage{amssymb}
\usepackage{amsmath}
\usepackage{epsfig}
\newcommand{\be}{\begin{equation}}
\newcommand{\ee}{\end{equation}}
\newcommand{\bea}{\begin{eqnarray}}
\newcommand{\eea}{\end{eqnarray}}

\newcommand{\D}{\displaystyle}
\newcommand{\g}{\gamma}
\newcommand{\f}{\frac}

\newcommand{\intc}[1]{{\int\frac{d#1}{2i\pi}}}
\newcommand\lr[1]{{\left({#1}\right)}}
\newcommand{\om}{\omega}

\begin{document}
\title{Next-leading BFKL effects in forward-jet production at HERA}
\author{O. Kepka}\email{oldrich.kepka@cea.fr}
\affiliation{DAPNIA/Service de physique des particules, CEA/Saclay, 91191 
Gif-sur-Yvette cedex, France}
\author{C. Marquet}\email{marquet@spht.saclay.cea.fr}
\affiliation{Service de physique th{\'e}orique, CEA/Saclay, 91191 
Gif-sur-Yvette 
cedex, France\\URA 2306, unit{\'e} de recherche associ{\'e}e au CNRS}
\author{R. Peschanski}\email{pesch@spht.saclay.cea.fr}
\affiliation{Service de physique th{\'e}orique, CEA/Saclay, 91191 
Gif-sur-Yvette 
cedex, France\\URA 2306, unit{\'e} de recherche associ{\'e}e au CNRS}
\author{C. Royon}\email{royon@hep.saclay.cea.fr}
\affiliation{DAPNIA/Service de physique des particules, CEA/Saclay, 91191 
Gif-sur-Yvette cedex, France}
%%%%%%%%%%%%%%%%%%%%%%%%%%%%%%%%%%%%%%%%%%%%%%%%%%%%%%%%%%%%%%%%%%
%%%%%%%%%%%%%%%%%%%%%%%%   Abstract   %%%%%%%%%%%%%%%%%%%%%%%%%%%%
%%%%%%%%%%%%%%%%%%%%%%%%%%%%%%%%%%%%%%%%%%%%%%%%%%%%%%%%%%%%%%%%%%
\begin{abstract}

We show that next-leading logarithmic (NLL) Balitsky-Fadin-Kuraev-Lipatov (BFKL) 
effects can be tested by the forward-jet cross sections recently measured at 
HERA. For $d\sigma/dx,$ the NLL corrections are small which confirms the 
stability of the BFKL description. The triple differential cross section 
$d\sigma/dxdk_T^2dQ^2$ is sensitive to NLL effects and opens the way for an 
experimental test of the full BFKL theoretical framework at NLL accuracy.

\end{abstract}
\maketitle

%%%%%%%%%%%%%%%%%%%%%%%%%%%%%%%%%%%%%%%%%%%%%%%%%%%%%%%%%%%%%%%%%%
%%%%%%%%%%%%%%%%%%%%%%%%   Section 1   %%%%%%%%%%%%%%%%%%%%%%%%%%%
%%%%%%%%%%%%%%%%%%%%%%%%%%%%%%%%%%%%%%%%%%%%%%%%%%%%%%%%%%%%%%%%%%

{\bf 1.}
Forward-jet production in deep inelastic scattering is a process in which a jet 
is detected at forward rapiditites in the direction of the proton. The 
virtuality of the intermediate photon $Q^2$ and the squared transverse 
momentum of the jet $k_T^2$ are two hard scales. When the total energy of the 
photon-proton collision $W$ is assumed to be large, corresponding to a small 
value of the Bjorken variable $x,$ forward-jet production is relevant 
\cite{mueller} for testing the Balitsky-Fadin-Kuraev-Lipatov (BFKL) approach 
\cite{bfkl}.

Indeed in the small$-x$ regime, besides the large logarithms coming from the 
strong ordering between the proton scale and the forward-jet scale (which are 
resummed using the Dokshitzer-Gribov-Lipatov-Altarelli-Parisi (DGLAP) 
evolution equation \cite{dglap}), other large logarithms arise in the hard 
cross section itself, due to the ordering between the energy $W$ and the 
hard scales. These can be resummed using the BFKL equation, at leading (LL) and 
next-leading (NLL) logarithmic accuracy \cite{bfkl,nllbfkl}. By contrast, in 
fixed-order perturbative QCD calculations the hard cross section is computed at 
fixed order with respect to $\alpha_s,$ and the next-to-leading order (NLO) 
predictions fail to describe the data.

NLL corrections to the LL-BFKL kernel were found to feature spurious 
singularities. It has been realised that renormalisation-group improved NLL 
regularisations can solve this singularity problem \cite{salam,CCS} and lead to 
consistent NLL-BFKL kernels. This, along with the success 
\cite{fjetsll,fjetsll2} of the LL-BFKL approach in describing the $d\sigma/dx$ 
data, motivates the present phenomenological analysis of NLL-BFKL 
effects in forward-jet production.

When fitting $d\sigma/dx,$ we obtain that the NLL corrections are small, which 
results in a good description of the H1 data by NLL-BFKL predictions. We also 
show that the recently measured triple differential cross section 
$d\sigma/dxdk_T^2dQ^2$ \cite{h1new} allows for a detailed study of the NLL-BFKL 
approach and of the QCD dynamics of forward jets. In particular, it has the 
potential to address the question of the remaining ambiguity corresponding to 
the dependence on the specific regularisation scheme of the NLL kernel. 

The present study is a phenomenological analysis of the new forward-jet 
data using NLL-BFKL predictions depending on the regularisation schemes and the 
renormalisation scale. In Ref. \cite{nllf2}, such a phenomenological 
investigation has been devoted to the proton structure function data, taking 
into account NLL-BFKL effects through an ``effective kernel'' (introduced in 
\cite{CCS}) using three different schemes. A saddle-point approximation for hard 
enough scales is used in order to obtain a phenomenological description of 
NLL-BFKL effects. In the present study devoted to forward-jet production, we 
implement them in a similar way. This allows to study the NLL-BFKL framework 
even though the determination of the next-leading impact factors is still in 
progress \cite{nllif}.

%%%%%%%%%%%%%%%%%%%%%%%%%%%%%%%%%%%%%%%%%%%%%%%%%%%%%%%%%%%%%%%%%%
%%%%%%%%%%%%%%%%%%%%%%%%   Section 2   %%%%%%%%%%%%%%%%%%%%%%%%%%%
%%%%%%%%%%%%%%%%%%%%%%%%%%%%%%%%%%%%%%%%%%%%%%%%%%%%%%%%%%%%%%%%%%

{\bf 2.} 
We shall use the usual kinematic variables of deep inelastic scattering: 
$x\!=\!Q^2/(Q^2\!+\!W^2)$ and $y\!=\!Q^2/(xs)$ where $\sqrt{s}$ is the 
center-of-mass energy of the lepton-proton collision. In addition, $x_J$ is the 
jet longitudinal momentum fraction with respect to the proton. The fully 
differential cross section for forward-jet production reads
\be
\f{d^{4}\sigma}{dxdQ^2dx_Jdk_T^2}=\f{\alpha_{em}}{\pi xQ^2}\sum_{\lambda=L,T}
\!f_\lambda(y)\ \f{d\sigma^{\g*p\!\rightarrow\!JX}_\lambda}{dx_Jdk_T^2}
\label{fj}\ee
with $f_T(y)\!=\!1\!-\!y\!+\!y^2/2,$ $f_L(y)\!=\!1\!-\!y,$ and 
where $d\sigma^{\g*p\!\rightarrow\!JX}_{T,L}/dx_Jdk_T^2$ is the cross section 
for forward-jet production in the collision of transversely (T) or 
longitudinally (L) polarized virtual photons with the proton.

In the following, we consider the high-energy regime $x\!\ll\!1$ in which the 
rapidity interval $Y\!=\!\log(x_J\!/\!x)$ is assumed to be large. Following 
the phenomenological NLL-BFKL analysis of \cite{nllf2}, one obtains 
for the forward-jet cross section:
\bea
\f{d\sigma^{\g*p\!\rightarrow\!JX}_{T,L}}{dx_Jdk_T^2}=
\f{\alpha_s(k_T^2)\alpha_s(Q^2)}{k_T^2Q^2}\ f_{eff}(x_J,k_T^2)\hspace{1.5cm}
\nonumber\\
\intc{\g}
\lr{\f{Q^2}{k_T^2}}^\g \phi^\g_{T,L}(\g)\ 
e^{\bar\alpha(k_T Q)\chi_{eff}\lr{\g,\bar\alpha(k_T Q)}Y}
\label{nll}
\eea
with the running coupling constant given by
\be
\bar\alpha(k^2)=\alpha_s(k^2)N_c/\pi=
\left[b\log\lr{k^2/\Lambda_{QCD}^2}\right]^{-1}
\label{runc}\ee
where $b\!=\!11/12\!-\!N_f/6N_c.$

In formula \eqref{nll}, $f_{eff}(x_J,k_T^2)\!=\!g\!+\!(q\!+\!\bar{q})C_F/N_c$
is the effective parton density which resums the leading logarithms in 
$\log(k_T^2/\Lambda_{QCD}^2).$ $g$ (resp. $q$, $\bar{q}$) is the gluon (resp. 
quark, antiquark) distribution function in the incident proton. Since the 
forward jet measurement involves large values of $k_T$ and moderate 
values of $x_J,$ formula \eqref{nll} features the collinear factorization of 
$f_{eff},$ with $k_T^2$ chosen as the factorization scale.

The NLL-BFKL effects are phenomenologically taken into account by the effective 
kernel $\chi_{eff}(\g,\bar\alpha).$ Indeed, the NLL-BFKL kernels
$\chi_{NLL}\lr{\g,\om}$ provided by the regularisation procedure obey a 
{\it consistency condition} \cite{salam,ccond} which allows to reformulate the problem in terms of $\chi_{eff}(\g,\bar\alpha).$ We shall consider the CCS scheme 
\cite{CCS} and the S3 and S4 schemes \cite{salam} in which $\chi_{NLL}$ also 
depends explicitly on $\bar\alpha.$ In each case, the effective kernel 
$\chi_{eff}(\g,\bar\alpha)$ is obtained from the NLL kernel 
$\chi_{NLL}\lr{\g,\omega}$ by solving the implicit equation
\be
\chi_{eff}=\chi_{NLL}\lr{\g,\bar\alpha\ \chi_{eff}}\ ,
\label{eff}
\ee
as a realization of the consistency condition.
\begin{figure}[t]
\begin{center}
\epsfig{file=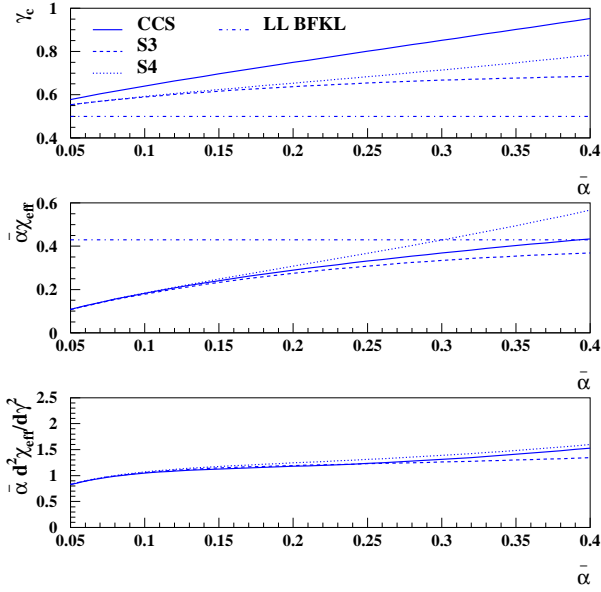,width=8cm}
\caption{$\g_c$, $\bar\alpha\chi_{eff}(\g_c,\bar\alpha)$ and 
$\bar\alpha\chi''_{eff}(\g_c,\bar\alpha),$ as functions of $\bar\alpha$ for the 
three BFKL resummation schemes CCS, S3 and S4. The fixed LL values are 
respectively $1/2,$ $0.43,$ and $5.47$ corresponding to $\bar\alpha=0.16.$}
\end{center}
\label{F2}
\end{figure} 

The details of the approximation \eqref{nll} are given in \cite{nllf2}, with the 
only difference that the kernel considered for $F_2$ was naturally the 
asymmetric one $\chi_{NLL}\lr{\g+\om/2,\om}.$ In the forward-jet problem, the 
energy scale is considered to be given symmetrically between the hard 
probes by $\log(W^2/k_TQ)$ instead of $\log(W^2/Q^2)\!\simeq\!\log(1/x)$ as was 
the case for $F_2.$ In other words, we do not perform the shift 
$\g\!\rightarrow\!\g\!+\!\om/2$ used for $F_2.$

Following Ref. \cite{renscal}, the renormalisation scale is $k^2\!\sim\!k_TQ.$ 
We have tested the sensitivity of our results when varying $k^2$ in the range 
$[k_TQ/\lambda,\lambda\ k_TQ]$ with $\lambda=2,$ with the substitution
$\bar\alpha(k_TQ)\!\rightarrow\!
\bar\alpha(\lambda k_TQ)\!+\!b\ \bar\alpha^2(k_TQ)\log(\lambda).$ Note that, 
following formula \eqref{eff}, the effective kernel is modified accordingly for 
each scheme, and we also modify the energy scale
$k_T Q\!\rightarrow\!\lambda\ k_TQ.$

It is important to note that in formula \eqref{nll}, we use the 
leading-order (Mellin-transformed) impact factors given by
\bea
\lr{\begin{array}{cc}
\phi^\g_{T}(\g)\\ \phi^\g_{L}(\g)
\end{array}}
=\pi\alpha_{em}N_c^2\sum_q 
e_q^2\f{1}{2\g^2}\lr{\begin{array}{cc}(1+\g)(2-\g)\\2\g(1-\g)\end{array}}
\nonumber\\\f{\Gamma^3(1+\g)\Gamma^3(1-\g)}
{\Gamma(2-2\g)\Gamma(2+2\g)(3-2\g)}
\label{phig}
\eea
as the full next-leading photon impact factors are not yet available (the jet impact factors are known at next-to-leading order \cite{ifnlo}). We point out that 
our phenomenological approach can be adapted to full NLL accuracy, once the 
next-leading impact factors available. 

Expressing the integral in \eqref{nll} using a saddle-point approximation in 
$\g,$ one finds for the theoretical forward-jet cross section
\bea
\f{d\sigma^{\g*p\!\rightarrow\!JX}_{T,L}}{dx_Jdk_T^2}\simeq
\f{\alpha_s(k_T^2)\alpha_s(Q^2)}{k_T^2Q^2}f_{eff}(x_J,k_T^2)
\phi^\g_{T,L}(\g_c)\nonumber\\
\lr{\f{Q^2}{k_T^2}}^{\g_c}e^{\D\bar\alpha\chi_{eff}(\g_c,\bar\alpha)Y}\ 
\f{\exp\lr{-\f{\log^2(Q^2/k_T^2)}{2\bar\alpha\chi_{eff}''(\g_c,\bar\alpha)\ 
Y}}}{\sqrt{2\pi\bar\alpha\chi_{eff}''(\g_c,\bar\alpha)\ Y}}\ ,
\label{nllsaddle}\eea where $\chi_{eff}''\!=\!d^2\chi_{eff}/d\g^2,$ and where 
the saddle point equation is $\chi_{eff}'(\g_c,\bar\alpha)\!=\!0.$ It is 
possible to extract the values of $\g_c,$ 
$\bar\alpha\chi_{eff}(\g_c,\bar\alpha)$ and 
$\bar\alpha\chi''_{eff}(\g_c,\bar\alpha)$ after solving the implicit equation 
\eqref{eff}. They are given in Fig.1 for the different schemes, as functions of 
$\bar\alpha.$

The description of the forward-jet cross section is then almost parameter free; 
the value of $\bar\alpha$ is imposed by the renormalisation group equations and 
only the overall normalisation is unknown (only the knowledge of the 
next-leading impact factors will provide a full prediction). By comparison, the 
LL-BFKL formula is formally the same as \eqref{nllsaddle}, but with the substitutions
\bea
\chi_{eff}&\rightarrow&\chi_{LL}(\g)= 2\psi(1)-\psi(1-\g)-\psi(\g)\nonumber\\
\g_c&\rightarrow&1/2\nonumber\\
\bar\alpha(k^2)&\rightarrow&\bar\alpha\!=\!const.\nonumber
\label{LL}\eea
with $\psi(\g)\!=\!d\log\Gamma(\g)/d\g.$ In the LL-BFKL case, this is a 
two-parameter formula (normalisation and $\bar\alpha$). The interesting property 
of our phenomenological approach is that formula \eqref{nll} has formally the 
structure of the LL formula, but with only one free (normalisation) parameter 
and a NLL kernel. The delicate aspect of the problem comes from the 
scheme-dependent effective kernel $\chi_{eff}.$

%%%%%%%%%%%%%%%%%%%%%%%%%%%%%%%%%%%%%%%%%%%%%%%%%%%%%%%%%%%%%%%%%%
%%%%%%%%%%%%%%%%%%%%%%%%   Section 3   %%%%%%%%%%%%%%%%%%%%%%%%%%%
%%%%%%%%%%%%%%%%%%%%%%%%%%%%%%%%%%%%%%%%%%%%%%%%%%%%%%%%%%%%%%%%%%

\begin{figure}[t]
\begin{center}
\epsfig{file=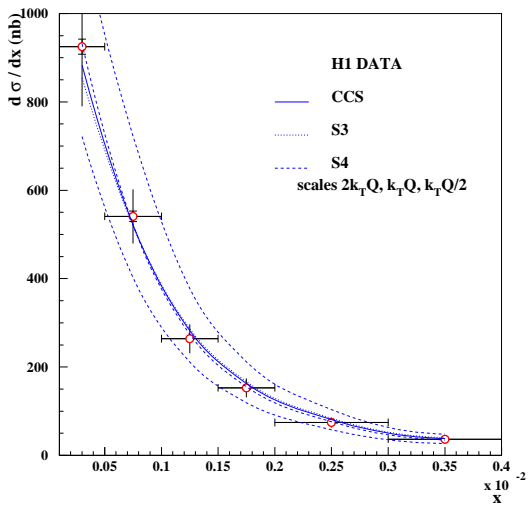,width=7cm}
\caption{The forward-jet cross section $d\sigma/dx$ measured by the H1 
collaboration, compared with the three NLL-BFKL parametrizations S4, CCS and 
S3 using the $k_TQ$ scale. For the S4 scheme, we also display the results
of the $2k_TQ$ and $k_TQ/2$ scales without changing the normalisation.}
\end{center}
\end{figure} 

{\bf 3.} The NLL-BFKL formula for the fully differential forward-jet cross 
section is obtained from \eqref{fj} and \eqref{nllsaddle}, with $\g_c$, 
$\bar\alpha\chi_{eff}(\g_c,\bar\alpha)$ and 
$\bar\alpha\chi''_{eff}(\g_c,\bar\alpha)$ shown in Fig.1 for the three 
different schemes.

To fix the normalisation (the only free parameter) and check the quality of
the data description using the BFKL formalism, we start by fitting the 
$d\sigma/dx$ H1 data \cite{h1new}. The choice of this data set 
corresponds to the kinematical domain where the BFKL formalism is expected to 
hold ($x\!\ll\!1$ and $Q^2/k_T^2\!\sim\!1$). The fitting procedure is the same
as the one described in Ref. \cite{fjetsll2}, Appendix A. The integrals over $x_J$, $Q^2$ and $k_T^2$ are performed numerically taking into account the kinematic cuts given in \cite{h1new}. We performed fits on statistical
errors only, systematics errors being strongly point-to-point correlated.
In other words, we do as if the systematic errors were 100\% 
correlated (which is close to reality). 

The fit results to the $d\sigma/dx$ H1 data are shown in Fig.2. The S4 fit can 
describe the data nicely ($\chi^2\!=\!5.4/5$ d.o.f.) whereas the S3 and CCS 
schemes show higher values of $\chi^2$ ($\chi^2\!=\!46.5/5$ and 
$\chi^2\!=\!22.2/5$ respectively). Note that, when fitting using total errors 
(shown on the figures), all three schemes give values of $\chi^2$ below 1. In 
any case, the fitted normalisations keep reasonable values, even with LL impact 
factors. We also display in Fig.2 the results obtained when varying the
renormalisation scale in the range $[k_TQ/\lambda,\lambda\ k_TQ]$ for the S4 scheme with $\lambda\!=\!2.$ We notice that this change of scale essentially affects the overall normalisation and thus does not alter the quality of the fit, after reajusting the normalisation. For larger values of $\lambda,$ the quality of the fit deteriorates, due to the high values reached by
$\bar\alpha.$ 

\begin{figure}[t]
\begin{center}
\epsfig{file=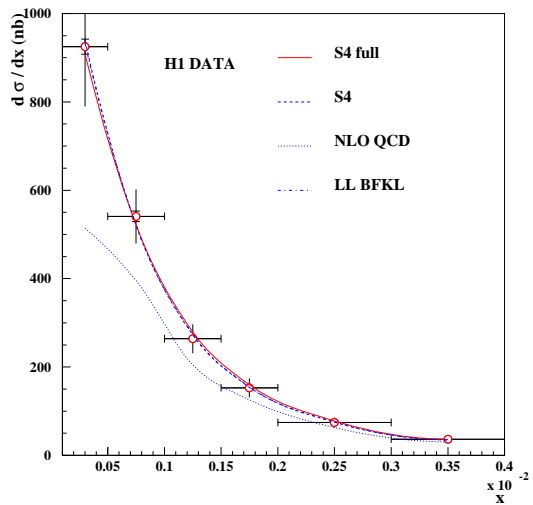,width=7cm}
\caption{The forward-jet cross section $d\sigma/dx$ measured by the H1 
collaboration, compared with LL and NLL (S4) BFKL fits and with NLO QCD 
predictions. The considered range of renormalisation scale for S4 does not alter 
the quality of the fit, after reajusting the normalisation.}
\end{center}
\end{figure}

A comparison of the S4 fit with the LL-BFKL results taken from Ref. \cite{fjetsll2} is shown in Fig.3. We notice the tiny difference between the LL and NLL results (the corresponding curves are not distinguishable on the figure). This confirms that the data are consistent with the BFKL enhancement towards small values of $x.$ Contrary to the proton structure function $F_2,$ the forward-jet cross section $d\sigma/dx$ does not show large NLL-BFKL corrections, once the overall normalisation fitted. This is due to the rather small value of the coupling $\bar\alpha\!\simeq\!0.16$ obtained in the LL-BFKL fit \cite{fjetsll2}, corresponding to an unphysically large effective scale.

We display in Fig.3 the result obtained when the $\gamma$ integration in \eqref{nll} is computed exactly, and it is compared with the one obtained when using the saddle-point approximation \eqref{nllsaddle}. Considering the moderate values of 
$Y\!=\!3-5$ probed by the forward-jet measurement, one could question the validity of the saddle-point approximation. In the case of the S4 scheme, the comparison with the exact computation shows that, after all the kinematic integrations have been performed, the difference can be absorbed in the overall normalization. This is also true for the S3 scheme. The case of the CCS scheme, for which there are difficulties to define the effective kernel in the whole complex plane, is left for future work.

We also present in Fig.3 the fixed order QCD calculation based on 
the DGLAP evolution of parton densities. The next-to-leading order (NLO) 
prediction of forward-jet cross sections is obtained using the NLOJET++ 
generator \cite{nlojetDIS}. CTEQ6.1M \cite{cteq} parton densities were used, and the renormalisation $\mu_r$ and the factorization scale $\mu_f$ were set equal to $\mu_r^2=\mu_f^2=Qk_t^{max},$ where $k_t^{max}$ corresponds to the maximal transverse momentum of forward jets in the event. The NLO QCD predictions do not describe the data at small values of $x,$ as they are below by a factor of order 2. 

The fit parameters obtained with statistical error only are used in the following to make predictions for other observables. Namely, the relative normalisations between the different NLL-BFKL calculations (CCS, S3 and S4) are used to make predictions for the triple differential cross section $d\sigma/dxdk_T^2dQ^2.$ This is an interesting observable as it has been measured with 3 different $k_T^2$ and $Q^2$ cuts, yielding 9 different regions for the ratio $r\!=\!k_T^2/Q^2.$ It was noticed in \cite{fjetsll2} that the LL-BFKL formalism leads to a good description of the data when $r$ is close to 1 and deviates from the data when $r$ is further away from 1, as effects due to the ordering between $Q$ and $k_T$ start to set in. NLO QCD predictions show the reverse trend.

\begin{figure}[t]
\begin{center}
\epsfig{file=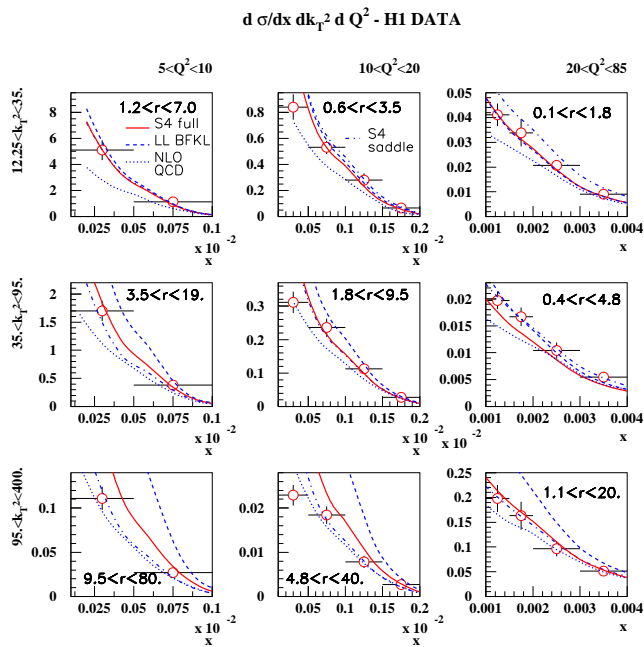,width=\columnwidth}
\caption{The forward-jet cross section $d\sigma/dxdk_T^2dQ^2$ measured by the H1 
collaboration, compared with LL and NLL (S4) BFKL fits and with NLO QCD predictions.
The S4 result with exact integration of \eqref{nll} is also shown, and it generally agrees with the saddle-point calculation.}
\end{center}
\end{figure} 

The H1 data for $d\sigma/dxdk_T^2dQ^2$ are shown in Fig.4 and compared with the 
S4 prediction, the LL-BFKL results (taken directly from \cite{fjetsll2}) and NLO QCD predictions. It is quite remarkable that the NLL-BFKL calculation, which includes some ordering between $Q$ and $k_T,$ leads to a good description of the H1 data on the full range. As it was the case for $d\sigma/dx,$ the difference between the LL and NLL results is small when $r\!\sim\!1.$ By contrast when $r$ differs from 1, the difference is significant, and the observable $d\sigma/dxdk_T^2dQ^2$ is quite sensitive to NLL-BFKL effects.

When varying the renormalisation scale in the range $[k_TQ/2,2k_TQ],$ the results shown in Fig.4 differ by a small amount \cite{tocome}, at most 10\%.
Also, we show in Fig.4 the curves obtained with the exact integration of 
\eqref{nll}, and one can see that they are quite close to the saddle-point results, except for large values of $r,$ where there is some deviation. Finally, we point out that the triple differential cross section has the potential to resolve the scheme ambiguity, as the predictions of the other schemes CCS and S3 do not compare with the data as well as the predictions of the S4 scheme \cite{tocome}.

%%%%%%%%%%%%%%%%%%%%%%%%%%%%%%%%%%%%%%%%%%%%%%%%%%%%%%%%%%%%%%%%%%
%%%%%%%%%%%%%%%%%%%%%%%%   Section 4   %%%%%%%%%%%%%%%%%%%%%%%%%%%
%%%%%%%%%%%%%%%%%%%%%%%%%%%%%%%%%%%%%%%%%%%%%%%%%%%%%%%%%%%%%%%%%%

{\bf 4.} Let us summarize our main results. For the cross section $d\sigma/dx,$ 
measured in the kinematical regime $Q^2/k_T^2\!\sim\!1,$ the difference between 
the LL-BFKL and NLL-BFKL descriptions is very small (see Fig.3), once the 
overall normalisation fitted. This confirms the validity of the BFKL description of \cite{fjetsll,fjetsll2} previously obtained with the LL formula and a (rather small) effective coupling. In the case of the triple differential cross section $d\sigma/dxdk_T^2dQ^2,$ the same conclusion holds when 
$r\!=\!k_T^2/Q^2\!\sim\!1.$ In addition when $r$ differs from 1, the NLL-BFKL 
description is quite different from the LL-BFKL one, as it is closer to the NLO 
QCD calculation (see Fig.3).

As a result, the best overall descrition of the 
data for $d\sigma/dxdk_T^2dQ^2$ is obtained with the NLL-BFKL formalism and the 
S4 scheme is favored. We need the complete knowledge of the next-leading impact 
factors before drawing final conclusions, in particular with respect to the 
overall normalisations, but our analysis strongly suggests that the data show 
the BFKL enhancement at small values of $x.$ This is of great interest in view 
of the LHC, where similar QCD dynamics will be tested with Mueller-Navelet jets 
\cite{mnjets,mnj}.

We would like to thank Laurent Schoeffel and Sebastian Sapeta for providing 
their numerical evaluations of the effective kernel.

%%%%%%%%%%%%%%%%%%%%%%%%%%%%%%%%%%%%%%%%%%%%%%%%%%%%%%%%%%%%%%%%%%
%%%%%%%%%%%%%%%%%%%%%%   Bibliography   %%%%%%%%%%%%%%%%%%%%%%%%%%
%%%%%%%%%%%%%%%%%%%%%%%%%%%%%%%%%%%%%%%%%%%%%%%%%%%%%%%%%%%%%%%%%%

\end{document}